\documentclass[conference]{IEEEtran}
\usepackage[ruled,vlined]{algorithm2e}

\ifCLASSINFOpdf

\else

\fi

\usepackage{amsfonts,epsfig,array,multirow,graphicx,amsmath,ltablex,tabularx,setspace,amssymb,multirow}
\usepackage{schemabloc,amsthm}
\usepackage{subfig}
\usepackage[numbers,sort&compress]{natbib}
\usetikzlibrary{circuits, arrows}
\usepackage{verbatim}
\usepackage{graphicx}
\usepackage{epstopdf}
\usepackage{mathtools}
\usepackage{xcolor}
\usepackage{marvosym}

\usepackage{lipsum}
\usepackage{dblfloatfix}
\usepackage{color}
\usepackage{etoolbox}
 \usepackage{epsfig}
\theoremstyle{plain}

\newtheorem{exm}{Example}

\makeatletter

\begin{document}
\SetKwRepeat{Do}{do}{while}
\newcolumntype{L}[1]{>{\raggedright\let\newline\\\arraybackslash\hspace{0pt}}m{#1}}
\newcolumntype{C}[1]{>{\centering\let\newline\\\arraybackslash\hspace{0pt}}m{#1}}
\newcolumntype{R}[1]{>{\raggedleft\let\newline\\\arraybackslash\hspace{0pt}}m{#1}}
\pgfdeclarelayer{background}
\pgfdeclarelayer{foreground}
\pgfsetlayers{background,main,foreground}
\newcommand{\oeq}{\mathrel{\text{\sqbox{$=$}}}}
\setlength{\textfloatsep}{0.1cm}
\setlength{\floatsep}{0.1cm}
\title{
\huge{Design of  Puncturing  for Length-Compatible Polar Codes Using Differential Evolution}
\vspace{-0.1in} }
\author{
   \IEEEauthorblockN{\small{Author-1 and Author-2 }}\vspace{-3.5em}}

\author{
    \IEEEauthorblockN{\small{ Kuntal Deka and Sanjeev Sharma}}}
   \author{ \small Kuntal Deka$^{1}$ and Sanjeev Sharma$^{2}$\\
   	$^1$ Indian Institute of Technology Goa,  India
   	$^2$Indian Institute of Technology (BHU) Varanasi, India \vspace{-0.2in}}
   
\vspace{-3.5in}
\maketitle
\begin{abstract}
 This paper presents a puncturing technique to design length-compatible polar codes. The punctured bits are identified with the help of differential evolution (DE).  A DE-based  optimization framework is developed where the  sum of the bit-error-rate (BER) values of the information bits is minimized.  We identify a set of bits which can  be avoided  for puncturing in the case of   additive white Gaussian noise (AWGN) channels. This reduces the size of the candidate puncturing patterns.  Simulation results confirm the superiority of the proposed technique over other state-of-the-art puncturing methods.
\end{abstract}


\IEEEpeerreviewmaketitle

\section{Introduction}
Polar code, proposed by Arikan \cite{arikan_2009}, is an important milestone in coding theory and has undoubtedly completed the long quest for   capacity-achieving codes.
In the original version \cite{arikan_2009}, the polarizing or the generator matrix was constructed by  the Kronecker power of the  binary  $2\times2$ kernel  ${\bf{F}}_2=\left[\begin{smallmatrix} 1&0\\ 1&1 \end{smallmatrix} \right]$. Due to this choice, the lengths are limited to powers of 2. Various polarizing kernels of larger size and defined over non-binary alphabets have been proposed \cite{nb_polar_rs,urbanke_polar_2010}.  However, these kernels  do not ensure low-complexity  decoding methods as in the case of ${\bf{F}}_2$.  Therefore, designing polar codes of arbitrary lengths with reasonable decoding complexity is a vital problem.

Puncturing is a simple and effective technique to modify the rate and the  length of a code. The rate of a polar code can be conveniently  adapted  by varying the number of frozen or information bits. Puncturing is not required for the rate-adaptability of a polar code. However, to attain length-compatibility for the polar codes, puncturing is very helpful. In \cite{QUP}, an efficient method is proposed  to design length-compatible polar codes. This method is referred to as the quasi-uniform puncturing (QUP). Suppose, one needs to puncture $n_p$ bits of a polar code of length $N$.  In QUP, the bit-reversed versions of the first $n_p$ consecutive integers $\left\{1,2, \cdots, n_p\right\}$ are considered for puncturing.    The  method in \cite{novel_puncturing}  selects the   bit-reversed versions of the last $n_p$ consecutive integers $\left\{N-n_p+1, \cdots, N\right\}$ as the puncturing bits.  The authors in \cite{shin_length_compatible} have proposed a puncturing technique by analyzing   the reduced polarization matrix after the  removal of the columns and the rows  corresponding to the punctured  and the frozen bits respectively.
In \cite{declercq_icc_2017}, the authors have partitioned the puncturing patterns into various equivalent classes and proposed a  method to find  the optimum pattern by  examining   only one representative of each class.
\subsection*{\underline{Contributions }}  In this paper,  the determination of the best puncturing pattern is formulated as an optimization problem.  Differential evolution (DE) is used for the optimization process.  DE is a popular and  simple evolutionary algorithm which is used  to solve complex optimization problems with real-valued parameters \cite{Storn1997}.  The suitability of various figures of merit or parameters for the objective function    is studied. After analyzing the behaviors of these parameters during the decoding process under puncturing, we decide to  consider the sum of the bit-error-rate (BER) values of the information bits   as the objective function.  The selection of the  information bits depends heavily on the puncturing pattern. We propose a DE-based search algorithm  to find the optimum  pair of the sets of the punctured and the information bits simultaneously  by minimizing the sum of the BER values for the information bits.  A technique to reduce the search-space for punctured bits  is presented where the even-indexed bits are overlooked.  

\section{Preliminaries}
Consider a polar code with the block length $N=2^m,~m\in \mathbb{Z}^+$.  The generator matrix is given by ${\bf{G}}_N={\bf{B}}_N{\bf{F}}_2^{\bigotimes m}$ where, ${\bf{B}}_N$ is the bit-reversal permutation matrix and $^{\bigotimes m}$ is the Kronecker power \cite{arikan_2009}.  For a  binary data vector $u_1^{{N}}=(u_{1}, u_{2},\ldots, u_{N})$, the  codeword ${x}^{N}_1$ is obtained by ${x}^{N}_1=u_1^{\textit{N}}{\bf{G}}_N$. This encoding process produces a set of $N$ polarized synthetic bit-channels.   For a rate $R=\frac{K}{N}$ code, the $K$ \textit{information} bits   are carried over the best $K$ bit-channels by putting them  into the respective slots $\cal{I}$ in $u_1^N$. The  bits in the other locations ${\cal{I}}^c$ are \textit{frozen} to 0 and these values are known perfectly to the decoder. The decoding is done by the successive cancellation (SC) algorithm \cite{arikan_2009}.

In order to derive a length-$N'$ polar code from a mother code of length $N$, a total of $n_p=N-N'$ bits of the codeword  $x_1^N$  need to be punctured. The rate of the modified code is given by $R'=\frac{K}{N-n_p}$.  Let ${\cal{P}}$ denote the set of puncturing bits with $\left|{\cal{P}}\right|=n_p$. The coded bits   corresponding to ${\cal{P}}$ are not transmitted.  The decoder knows only the location of the punctured bits and sets their initial log-likelihood ratio (LLR) values to zero. Because of the puncturing of the bits in $\cal{P}$, the quality of the synthesized  bit-channels get modified and the information set $\cal{I}$ should  be re-selected.

\section{Design of Puncturing Pattern  based on Differential Evolution}

Suppose the objective is to derive a  length-$N'$ polar code from a length-$N$ one. For that, one needs to puncture $n_p=N-N'$ bits. The number of candidate bits is $D=N$ and we have to select the best $n_p$ bits amongst these $D$ bits.  
The optimization problem can be formulated as:
\begin{equation}
\label{opt_problem}
\left[{\cal{P}}_m, {\cal{I}}_m\right] =\arg \min_{{\cal{P}}, {\cal{I}}} f\left({\cal{P}}, {\cal{I}}, \frac{E_b}{N_0}\right)
\end{equation}
where, the objective function is $ f\left({\cal{P}}, {\cal{I}}, \frac{E_b}{N_0}\right)$ and $\frac{E_b}{N_0}$ is the signal-to-noise-ratio (SNR).  There are many figures of merit which can be considered as the objective function. Some of these are Bhattacharrya parameters of the bit-channels, the BER values of the individual bits computed by Monte Carlo simulation, the mean of the LLRs etc.  These parameters are also taken into consideration in the construction step \cite{harish_construction}.
In order to find the best figure of merit for puncturing, we analyze the evolution of various parameters during decoding under the influence of puncturing. 
 \begin{figure}[!ht] 	
 	\centering
 	\subfloat[$x_1$ punctured]{\includegraphics[height=3cm, width=4.2cm]{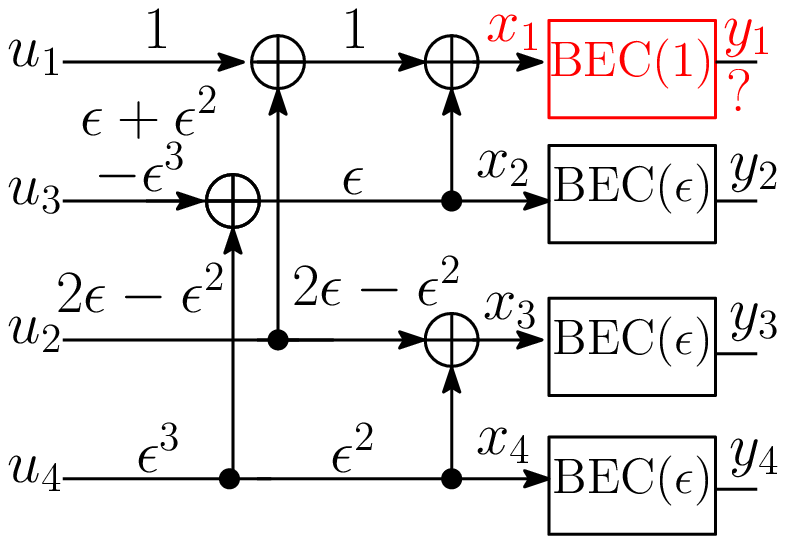} \label{BEC_x1} }	
 	\hspace{0.001in}
 	\subfloat[$x_4$ punctured]{\includegraphics[height=3.11cm, width=4.2cm]{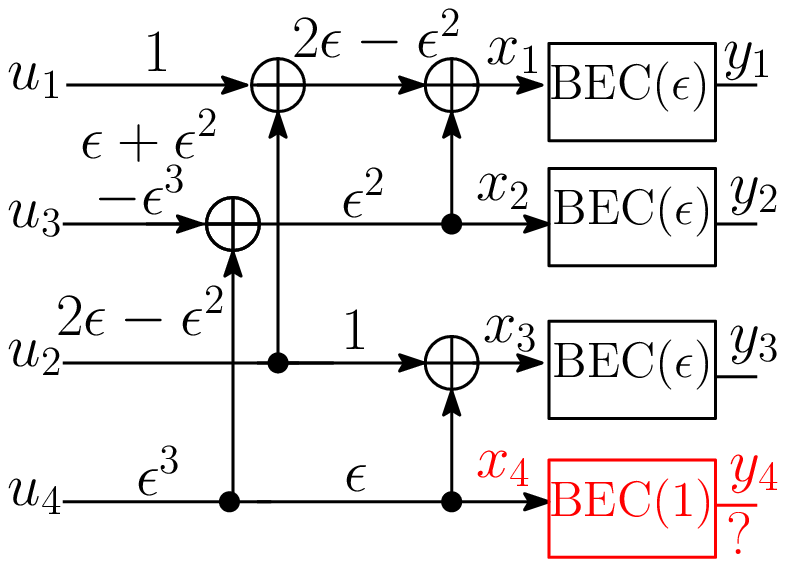}\label{BEC_x2}}

 	\caption{\footnotesize{Bhattacharyya parameters for $N'=3$ polar code over BEC with erasure probability $\epsilon$.}}
 	\label{pu1}
 \end{figure}
 %
 
 Consider the  generation of $N'=3$ polar code from $N=4$ mother polar code by puncturing one bit in the case of binary erasure channel (BEC). In Fig.~\ref{pu1}(a), the coded-bit $x_1$ is punctured. Since, this bit is completely erased,  the first channel effectively becomes a  BEC with erasure probability 1. The other channels are identical  and equal to BEC with erasure probability $\epsilon$. 
 By applying Proposition~6 of \cite{arikan_2009}, the evolution  of these parameters at different layers is shown in Fig.~\ref{pu1}(a) when $x_1$ is punctured. These are  found to be  $\left\{1, 2\epsilon-\epsilon^2,\epsilon+\epsilon^2-\epsilon^3, \epsilon^3 \right\}$ for the bit-channels.  Consider the case when $x_4$ is punctured instead of $x_1$ as shown in Fig.~\ref{pu1}(b). The Bhattacharyya parameters are the same as that in the previous case. 
  \begin{figure}[!htbp] 	
 	\centering
 	\subfloat[$x_1$ punctured]{\includegraphics[height=3cm, width=4.3cm]{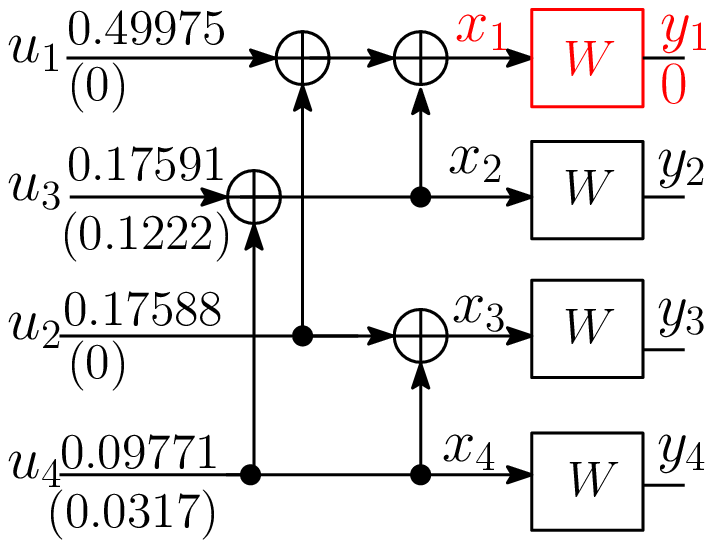}}
 	\hspace{0.003in}
 	\subfloat[$x_4$ punctured]{\includegraphics[height=3cm, width=4.3cm]{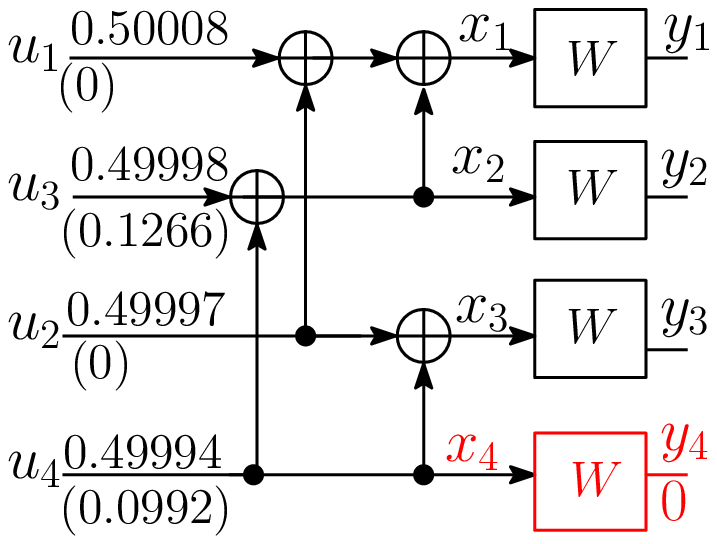}}
 	\caption{\footnotesize{BER values for $N'=3$ polar code over AWGN channel at 1 dB. }}
 	\label{pu3}
 \end{figure}
 This means that the puncturing patterns $\left\{1\right\}$  and $\left\{4\right\}$ are equivalent when the underlying channel is BEC.  However, for other channels, these two puncturing patterns may not be equivalent. Fig.~\ref{pu3}  shows such a situation when the underlying channel is AWGN (represented by $W$). The BER values of the input bits as computed from Monte Carlo simulation are $\{0.49975, 0.17588, 0.17591, 0.09771\}$ and $\left\{0.50008, 0.49997, 0.49998, 0.49994\right\}$ for the puncturing patterns $\{1\}$ and $\{4\}$ respectively at $\frac{E_b}{N_0}=1$ dB\footnote{  The Monte Carlo method in \cite{arikan_2009} was used to find the estimates for the Bhattacharyya parameters of the bit channels. As these parameters are related to the probability of error for the input bits,  we consider the Monte Carlo simulation to estimate the probability of bit error.}.  This shows that $\{1\}$ is better than $\{4\}$  and in fact $\{4\}$  should be avoided for puncturing.
 Observe that here, we have considered  all the input bits to be the  information bits.

 %
 The  BER values  after the selection of the information bits are also analyzed here.   These BER values are more appropriate measures and are shown within brackets in Fig.~\ref{pu3} when the rate $R=0.5$. 
 	It can be safely  concluded that $\{x_1\}$ is a better puncturing pattern than $\{x_4\}$. 
 \begin{algorithm}[!htbp]
 	\caption{\small Puncturing based on differential evolution }
 	\label{algo:puncturing_DE}
 	\small
 	\SetKwData{Left}{left}
 	\SetKwData{This}{this}
 	\SetKwData{Up}{up}
 	\SetKwFunction{Union}{Union}
 	\SetKwFunction{FindCompress}{FindCompress}
 	\SetKwInOut{Input}{input}
 	\SetKwInOut{Output}{output}
 	\Input{$N$, $K$, $n_p$,  $\frac{E_b}{N_0}$,  $F$,  $C_r$, and \textcolor{black}{$S_P$}		}
 	\Output{Puncturing bits ${\cal{P}}_m$ and information bits ${\cal{I}}_m$}
 	\BlankLine
 	%
 	Initialize the population matrix ${\bf{P}}$ to a  matrix of size $S_P \times D$ having random numbers uniformly distributed over $[0,1]$ where $D=N$  \tcp*[f]{\scriptsize{Initialization}}\;
 	
 	\While{termination criteria not fulfilled}{
 		\For{$i\leftarrow 1$ \KwTo $S_P$}{	
 			\textcolor{black}{Select three distinct vectors (rows) ${\bf{z}}_{r_0}$,${\bf{z}}_{r_1}$ \text{ and } ${\bf{z}}_{r_2}$ uniformly at random from $\bf{P}$ such  that they are also different from ${\bf{z}}_i$}\;
 			Generate an integer $j_{\text{rand}}$ uniformly at  random from $\left\{1,2, \ldots, D \right\}$\;
 			\tcc{\scriptsize{Generation of trial vector ${\bf{u}}$}}
 			\For{$j\leftarrow 1$ \KwTo $D$}{
 				\If{${\text{rand[0,1]}} \leq C_r$  or $j=j_{\text{rand}}$ }{
 					$\textcolor{black}{{\bf{w}}_{j,i}}={\bf{z}}_{j,r_0}+F\times\left({\bf{z}}_{j,r_1}-{\bf{z}}_{j,r_2}\right)$ \tcp*[f]{\scriptsize Crossover and Mutation}
 					
 				}
 				\Else{
 					${\bf{w}}_{j,i}={\bf{z}}_{j,i}$
 				}	
 				
 			}	
 			\tcc{\scriptsize{Evaluation and Selection}}	
 			Suppose ${\bf{w}}^{p}$ and ${\bf{z}}_i^{p}$ are the first $n_p$ arguments/indices of the  sorted (descending) version of
 			${\bf{w}}$ and   ${\bf{z}}_i$ respectively.
 			
 			With the help of GA method, find the sets ${\cal{I}}_{{\bf{w}}^{p}}$ and ${\cal{I}}_{{\bf{z}}_i^{p}}$ of the  information bits when the code bits in ${\bf{w}}^{p}$ and ${\bf{z}}^{p}_i$  are punctured respectively\;
 			Run Monte Carlo simulation with the chosen information or frozen sets.  Suppose,  $f({\bf{w}}^{p})$ and $f({\bf{z}}^{p}_i)$ are the sums of the BER values  for the information bits in ${\cal{I}}_{{\bf{w}}^{p}}$ and  ${\cal{I}}_{{\bf{z}}_i^{p}}$ respectively  \;
 			\If{$f({\bf{w}}^{p}) <f({\bf{z}}^{p}_i)$}{
 				${\bf{z}}_i={\bf{w}}$  \tcp*[f]{\scriptsize{Replace the $i$th row of $\bf{P}$ by $\bf{w}$}}\;
 			}		
 		}
 		From the updated population matrix $\bf{P}$, find the vector (row) ${\bf{z}}_{\min}$ (or equivalently ${\bf{z}}_{\min}^{p}$ ) which yields the minimum value  of objective function\;
 		If $f\left({\bf{z}}_{\min}^{p}\right)$ is not changing significantly from the previous iteration or the maximum number of iterations are exhausted, then break  from  loop\;
 	}
 	
 	Set ${\cal{P}}_m$ to ${\bf{z}}_{\min}^{p}$ \;
 	Assign the  set ${\cal{I}}_{{\bf{z}}_{\min}^{p}}$ of information bits to ${\cal{I}}_m$\;

 \end{algorithm}
The above examples show that the Bhattacharyya parameters  are not suitable for designing puncturing patterns for general channels. The BER values computed from Monte Carlo simulation are more reliable features. Therefore, in (\ref{opt_problem}), we consider the objective function   $ f\left({\cal{P}}, {\cal{I}}, \frac{E_b}{N_0}\right)$ as the sum of the BER values of the  bits in the information set $\cal{I}$ at SNR $\frac{E_b}{N_0}$ when the coded bits in $\cal{P}$ are punctured. For brevity, $ f\left({\cal{P}}, {\cal{I}}, \frac{E_b}{N_0}\right)$ will be substituted by $ f\left({\cal{P}}\right)$ with the understanding that $\cal{I}$  is the optimum information  set for  ${\cal{P}}$ at a fixed SNR$=\frac{E_b}{N_0}$.

In order to solve (\ref{opt_problem}), we adopt  DE.  
The detailed steps are shown in Algorithm~\ref{algo:puncturing_DE}. In DE,  a population $\bf{P}$ of vectors is updated  iteratively.  The number of vectors in the population is denoted by $S_P$.  The length of a vector is $D=N$ in this case. At first,   $\bf{P}$ is initialized as a matrix of dimension $S_P \times D$ whose elements are chosen uniformly at random from $[0, 1]$.
For each vector ${\bf{z}}_i$ , $i =1,\ldots, S_P$ in $\bf{P}$, a trial vector $\bf{w}$ is generated with the given values of crossover rate ($C_r$) and scaling factor ($F$). The details of the generation of the trial vector are presented in  Algorithm~\ref{algo:puncturing_DE}.   Here, we consider the convention that, for any candidate vector ${\bf{z}}_{i}=\left({\bf{z}}_{1,i}, \ldots, {\bf{z}}_{D,i}\right)$, if ${\bf{z}}_{j,i}>{\bf{z}}_{k,i}$, then it is preferable to puncture the $j$th bit compared to the $k$th bit as per that candidate. Based on this convention, the vectors $\bf{w}$ and ${\bf{z}}_i$ are sorted in descending order and the arguments are stored in ${\bf{w}}^{\text{sorted,arg}}$ and ${\bf{z}}_i^{\text{sorted,arg}}$ respectively.  Then the first  $n_p$ indices are stored in ${\bf{w}}^{p}$  and ${\bf{z}}_i^{p}$.  By using Gaussian approximation (GA) method \cite{trifonov}, the information bits ${\cal{I}}_{{\bf{w}}^{p}}$ and ${\cal{I}}_{{\bf{z}}_i^{p}}$  are found out against the puncturing patterns ${\bf{w}}^{p}$ and  ${\bf{z}}_i^{p}$ respectively.   Note that the  information bits need to be re-selected  for every distinct puncturing pattern. GA is considered for the construction step as it  provides good performance with low complexity \cite{harish_construction}. Now, by carrying out Monte Carlo simulation, the values of the objective functions $f({\bf{w}}^{p})$ and $f({\bf{z}}^{p}_i)$ are computed. If $f({\bf{w}}^{p})<f({\bf{z}}^{p}_i)$, then the $i$th row of $\bf{P}$ is replaced by the trial vector $\bf{w}$.  In this way,  every vector in $\bf{P}$  is examined and updated if needed.  From the updated ${\bf{P}}$, the best vector ${\bf{z}}_{\min}^{p}$   with the minimum objective value is found out. If  there is negligible change in this objective value from the previous iteration or the maximum number of iterations are completed, the algorithm is stopped. The puncturing pattern ${\bf{z}}_{\min}^{p}$ and the corresponding set of the information bits ${\cal{I}}_{{\bf{z}}_{\min}^{p}}$ are returned as the outputs ${\cal{P}}_m$  and ${\cal{I}}_m$.

\begin{figure}[!ht] 	
	\centering
	\includegraphics[height=2cm, width=8cm]{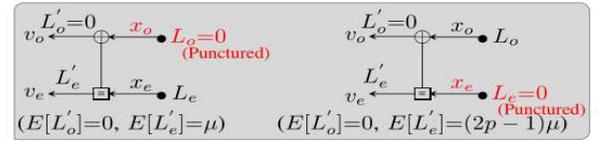}
	\caption{\footnotesize{LLR-based SC decoding under puncturing at the last layer ($v_o$  and $v_e$ are typically intermediate bits and not the frozen/information bits).}}
	\label{LLR}
\end{figure}

{\textbf{\textit{\textcolor{black}{Reduction of the search space}}}}: \textit{
	For length-compatible  polar codes,  the search space for the punctured bits can be reduced by ignoring the  set $F_P$ of forbidden bits as  given by 
		$$
		F_P={\cal{E}}\cup \left\{N-1\right\}
		$$
		where, ${\cal{E}}=\left\{2,4, \cdots, N-2, N\right\}$ is the set of even-indexed bits.
		Polar codes of any arbitrary length can be obtained without resorting to  puncturing of  these forbidden bits. We set $D=\frac{N}{2}-1$ in Algorithm~\ref{algo:puncturing_DE}.}\\
{\textbf{\textit{\textcolor{black}{Justification}}}}:
The polar encoding structure for length-$N$ code contains $\log_2 N$ layers with each layer containing $N/2$ basic butterfly structures. The structure contains $N$ branches corresponding to the coded bits.  
\begin{figure}[!ht] 	
	\centering
	\includegraphics[height=5.7cm, width=6.7cm]{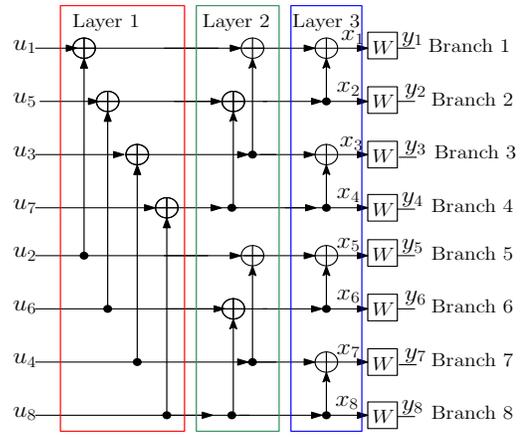}
	\caption{\footnotesize{Encoding structure for $N = 8$ polar code.}}
	\label{p8}
\end{figure}
The situation is explained in Fig.~\ref{p8} for the case $N=8$. 
The input bits comprising of the frozen and the information bits are fed to the first layer.  The last layer is connected directly to the channels.  
Consider the  SC decoding in LLR domain  over a particular basic structure in the last layer  as shown in Fig.~\ref{LLR}.	

   The input LLRs to the upper (odd) and the lower (even) branch of the basic structure are $L_o$ and $L_e$ respectively. The  outputs  are given by:
\vspace{-0.1in}
\begin{equation}
	\begin{aligned}
		L_o'=& 2\tanh^{-1}\left[\tanh\left(\frac{L_o}{2}\right)\tanh\left(\frac{L_e}{2}\right)\right] \\
		L_e'=& (1-2\hat{v}_o)L_o +L_e
	\end{aligned}
	\label{SC_LLR}
\end{equation}
where, $\hat{v}_o$ is the most recent  estimate found regarding the bit $v_o$ while computing $L_e'$.

We take insight from  the GA method where the mean of the LLR messages is updated across the layers \cite{trifonov}. Consider the transmission of all-zero codeword. Suppose, $\mu$ is the mean of the channel LLR values.  
As shown in Fig.~\ref{LLR}, if the upper or the odd bit $x_o$ is punctured, then $L_o=0$.  Subsequently, by (\ref{SC_LLR}), the output LLRs become $L_o'=0, L_e'=L_e$.  Thus we have the following pair of mean values $\left(E\left[L_o'\right]=0, E\left[L_e'\right]=\mu \right)$. 
On the other hand, if the lower or the even bit $x_e$ is punctured, then $L_e=0$. In that case, $L_o'=0$. Note  that  the computation of   $\hat{v}_o$  may benefit from the known values of a few frozen bits by the time $L_e'$ is computed.   Suppose, $p$ is the probability that $\hat{v}_o$  is correct i.e., $\Pr\left(\hat{v}_o=v_o=0\right)=p$.  In that case, the pair of mean values are given by $\left(E\left[L_o'\right]=0, E\left[L_e'\right]=(2p-1)\mu \right)$.  As $p\leq 1$, we have $(2p-1)\mu \leq \mu$.   Therefore, when  $x_e$ is punctured, the evolution of the mean is  slower compared to case where  $x_o$ is punctured.
\begin{table}[h!]	
	\caption{Number of appearances}
	\centering
	\scalebox{0.8}{
		\begin{tabular}{|C{3cm}|c|c|c|c|c|c|c|c|} 
			\hline
			\textbf{Bit/ Branch} & $x_1$ & $x_2$  & $x_3$	 & $x_4$  & $x_5$  & $x_6$  & $x_7$  & $x_8$\\ \hline
			\textbf{As upper branch} & 3 &  2 & 2 & 1 & 2 & 1 & 1 & 0 \\ \hline
			\textbf{As lower branch}& 0 & 1 & 1 & 2 & 1 & 2 & 2 & 3 \\ \hline		
		\end{tabular}}
		\label{involvement}
	\end{table}
 In GA, the probability of bit error  is inversely proportional to the mean value.  This implies that the probability of error for the information bits will be higher when $x_e$ is punctured.  Moreover, both the upper and the lower bits of a basic structure should not be punctured simultaneously because it will fully disturb the structure.   Therefore, the search space may be reduced by rejecting all even bits $\left({\cal{E}}=\left\{2,4,\ldots, N\right\}\right)$. The total number of even bits is $N/2$.  The maximum number of bits to be punctured is $N/2-1$. Amongst the odd bits, the bit  or branch`$N-1$' appears as the lower branch in the maximum number of basic structures  in various layers. The number of involvements of a bit as lower and upper branch  are shown in TABLE~\ref{involvement}. 
The set ${F}_P$ of  the forbidden bits is given by $F_P={\cal{E}}\cup \left\{N-1\right\}$. There is no need to puncture any of the bits in $F_P$ as any lower-length code can be derived from a  code of length $N/2$ or less.  $\blacksquare$

\begin{exm}			
		Consider the case of deriving $N'=6$ polar code from $N=8$ polar code by puncturing $n_p=2$ coded bits.  The DE-based algorithm is invoked to find the best $n_p=2$ bits for puncturing.   We consider a population matrix  $\bf{P}$ of size  $4\times3$ with $S_P=4$ and $D=\frac{8}{2}-1=3$.   ${\bf{P}}$ is initialized to a random matrix where an element is selected uniformly at random from [0,1]. Suppose ${\bf{P}}$ is initialized to  the following matrix:
		\begin{equation}
		\label{populationmatrix2}
		{\bf{P}}=\left[ {\begin{array}{ccc}
			0.68471631 & 0.144816 &  0.26360207\\
			0.0790236 & 0.40264467 & 0.13473581\\
			0.59553136 & 0.57930957 & 0.77943687 \\
			0.96593194 & 0.03113405 &  0.83083448
			\end{array} }\right]. \end{equation}	
		For every row of $\bf{P}$, a trial vector is generated by carrying out the  mutation and the crossover operations.  For the selection step, we
		consider the sum of the BER values of the information bits as the objective function.    The punctured bits are  identified from  the indices of the  sorted rows of $\bf{P}$. The first column refers to puncturing of  bit 1, the second column refers to puncturing of bit 3  and the third column refers to puncturing of bit 5. For example, consider the first row $\left(	0.68471631,  0.144816,  0.26360207\right)$ of $\bf{P}$ in (\ref{populationmatrix2}).   As we need to select two bits for puncturing, we consider the indices of the first two highest row elements. The first two highest elements are $\left(0.68471631,0.26360207\right)$ and they refer to puncturing of  $\left(1,5\right)$. For these punctured bits, the information bits are selected using GA.  Monte Carlo simulation for SC decoding is carried out. The sum of the BER values of the information bits  is considered as the objective function during the selection process. If the value of objective function for the first row is higher than that for the trial vector, then the first row is replaced by the trial vector.   In this way, every row of   $\bf{P}$ is examined and  updated iteratively if required. When the stopping criteria are met,  the best row or vector  (having  the lowest sum of the  BER values) from $\bf{P}$ is selected and the corresponding set of punctured bits is considered as the optimum  pattern.
\end{exm}

 \section{Simulation Results}
 In recent communication standards, polar codes of short blocklengths have been considered \cite{5G_NR}. 
 We present the simulation results for two cases.
 The short codes are  considered  so that the punctured bits and the information bits can be explicitly mentioned. 
   Due to space constraint, we provide only the block-error-rate (BLER) performances  although the BER results are found to be equally impressive.

{\textbf{\textit{\underline{Case 1}}}}:
In this case, we  puncture $n_p=28$ bits of   polar code of length $N=128$ and rate  $R=0.5$.  This puncturing will produce a  code of  length $N'=100$ and rate  $R'=0.64$.   The DE-based algorithm is run  to find the optimum punctured bits and information bits with the parameters $S_P=100, C_r=0.8$ and $F=0.6$ at  $\frac{E_b}{N_0}=6$ dB. These bits are shown in Table~\ref{tab:table4}.  The DE-based search algorithm is run to find the optimum puncturing pattern at  an SNR such that the BER is around $10^{-5}$. The pattern  determined  in this way is found  to work well at different SNR values.
\begin{table}[!htbp]
	\begin{center}
		\caption{\footnotesize{${\cal{P}}_m$  and ${\cal{I}}_m$ for  Case 1}}
		\label{tab:table4}
		\scalebox{0.75}{
			\begin{tabular}{|L{2.6cm}|L{7cm}|} 
				\hline
				\textbf{Punctured bits  ${\cal{P}}_m$} & 1     3     5     7      9    11    13    17    21    25    33    37    41    45    49    53    57    65    69    73    77    81  85    89    97   101   105    113 				\\ \hline 
				\textbf{Information bits ${\cal{I}}_m$} & 32 46 47 48 52 54 55 56 58 59 60 61 62 63 64 72 76 78 79 80 84 85 86 87 88 89 90 91 92 93 94 95 96 98 99 100 101 102 103 104 105 106 107 108 109 110 111 112 113 114 115 116 117 118 119 120 121 122 123 124 125 126 127 128
				\\
				\hline 			
			\end{tabular}}
		\end{center}
	\end{table}
\begin{figure}[!ht] 	
	\centering
	\includegraphics[height=6.5cm, width=8cm]{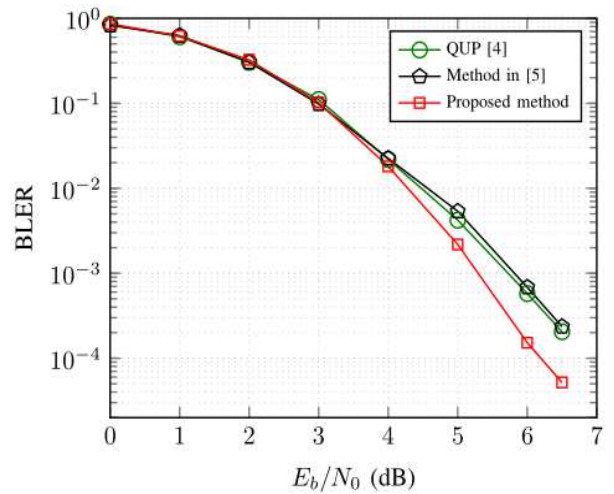}
	\caption{\footnotesize{Comparison  under  SC  decoding, Case 1.}}
	\label{case4}
\end{figure}

		
	The BLER performances of the puncturing methods  under SC decoding are shown in Figure~\ref{case4}.   Observe that the proposed puncturing pattern yields the best
	result and offers a coding gain of about 0.8~dB at BLER=$10^{-4}$.    The high value of the coding gain confirms the superiority of the DE-based puncturing strategy over the existing methods.    

\begin{figure}[!ht] 	
	\centering
	\includegraphics[height=6.5cm, width=8cm]{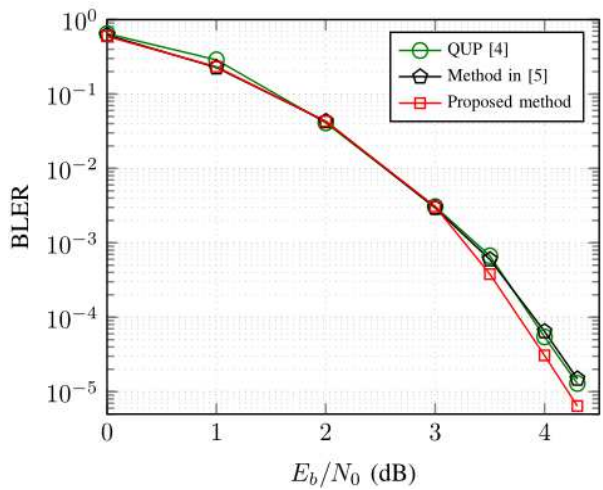}
	\caption{\footnotesize{Comparison  under CRC-aided SC list decoding, Case~1.}}
	\label{list_decoding_case4}
\end{figure}
 	
	
We also evaluate the performances of these puncturing
schemes under cyclic-redundancy-check (CRC) aided SC list
decoding \cite{list_decoding}. The size of a list is set to $L = 8$. We consider an
outer CRC code of length 16 with generator polynomial   $g(x) = x^{16} + x^{12} +
x^5 + 1$. This code is known as CRC-16-CCITT.	   The CRC coded bits are put in the locations of the   last 16 
	 information bits as per the recommendation given in \cite{list_decoding}.   The performances of  the puncturing
	schemes under CRC-aided SC list
	decoding are shown in Figure~\ref{list_decoding_case4}.  Observe that,  the proposed puncturing method performs better than the QUP \cite{QUP} and method in \cite{novel_puncturing}. 
	However, unlike in the case of SC
	decoding, the coding gain is relatively small and it is around 0.25~dB at BLER=$10^{-4}$.	
This reduction of the coding gain is due to the presence  of a powerful CRC code as the outer code in the concatenated encoding scheme.  Nevertheless, the proposed puncturing method performs significantly better than the existing methods in a purely polar coding environment.

{\textbf{\textit{\underline{Case 2}}}}:
In this case, we  puncture $n_p=24$ bits of a  polar code of length $N=64$ and rate  $R=0.5$.  This puncturing will produce a  code of  length $N'=40$ and rate  $R'=0.8$.   The DE-based algorithm is run  to find the optimum punctured bits and information bits with $S_P=50, C_r=0.8$ and $F=0.6$ at  $\frac{E_b}{N_0}=8$ dB. These bits are shown in Table~\ref{tab:table3}.  
	\vspace{-0.1in}
\begin{table}[h!]
	\begin{center}
		\caption{\footnotesize{${\cal{P}}_m$  and ${\cal{I}}_m$ for  Case 2}}
		\label{tab:table3}
		\scalebox{0.75}{
			\begin{tabular}{|L{2.6cm}|L{7cm}|} 
				\hline
				\textbf{Punctured bits  ${\cal{P}}_m$} & 1,  3,  5,  7,  9,  11,  13,  15,  17,  19,  21,  23,  27,  29,  33,  37,  39,  41,  45,  51,  53,  55,  59,  61	\\ \hline
				\textbf{Information bits ${\cal{I}}_m$} & 24 28 30 31 32 36 38 39 40 42 43 44 45 46 47 48 49 50 51 52 53 54 55 56
				57 58 59 60 61 62 63 64
				\\
				\hline 			
			\end{tabular}}
		\end{center}
	\end{table}
	\begin{figure}[!ht] 	
		\centering
		\includegraphics[height=6.5cm, width=8cm]{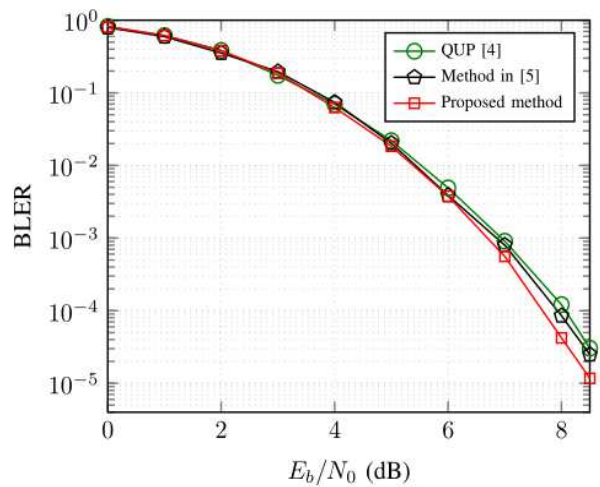}
		\caption{\footnotesize{Comparison  under  SC  decoding, Case 2.}}
		\label{case3}
	\end{figure}
    The BLER performances of the puncturing methods  under SC decoding are shown in Figure~\ref{case3}.   Observe that the proposed puncturing pattern yields the best
	result and offers a coding gain of about 0.3~dB at BLER=$10^{-4}$.    This coding gain is smaller than that in the previous case.  This is due to the fact that a higher number of bits are  punctured which, in turn, produces  a code with a  high rate of $R'=0.8$. 

	\begin{figure}[!ht] 	
		\centering
		\includegraphics[height=6.5cm, width=8cm]{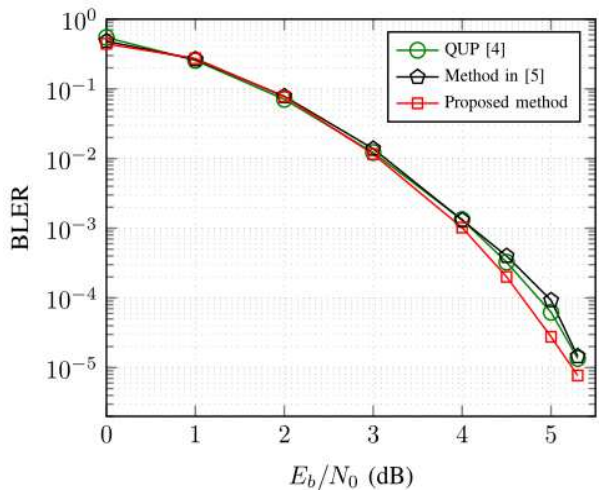}
		\caption{\footnotesize{Comparison under CRC-aided SC list decoding, Case~2.}}
		\label{list_decoding_case3}
	\end{figure}	
	
    The performances of  the puncturing
	schemes under CRC-aided SC list
	decoding are shown in Figure~\ref{list_decoding_case3}.  The CRC coded bits are put in the locations of the last 16
	information bits. Observe that, in this case also, the proposed puncturing method performs better than the QUP \cite{QUP} and the method in \cite{novel_puncturing}.  Similar to the previous case, we have experienced  a reduction in the coding gain. The coding gain is  around 0.2~dB at BLER=$10^{-4}$.

\section{Conclusions}
This paper presented a DE-based technique to search for the optimum  pair of the puncturing and the  information bits for length-compatible polar codes. By analyzing the decoding progression under puncturing, the even-indexed bits and the last odd-indexed bit are excluded from the search space. DE-based optimization is carried over this reduced space. Simulation results are provided  to  compare the proposed method with other methods in literature.

\bibliographystyle{ieeetran}
\footnotesize
\bibliography{references_polar}

\begin{thebibliography}{10}
\providecommand{\url}[1]{#1}
\csname url@samestyle\endcsname
\providecommand{\newblock}{\relax}
\providecommand{\bibinfo}[2]{#2}
\providecommand{\BIBentrySTDinterwordspacing}{\spaceskip=0pt\relax}
\providecommand{\BIBentryALTinterwordstretchfactor}{4}
\providecommand{\BIBentryALTinterwordspacing}{\spaceskip=\fontdimen2\font plus
\BIBentryALTinterwordstretchfactor\fontdimen3\font minus
  \fontdimen4\font\relax}
\providecommand{\BIBforeignlanguage}[2]{{%
\expandafter\ifx\csname l@#1\endcsname\relax
\typeout{** WARNING: IEEEtran.bst: No hyphenation pattern has been}%
\typeout{** loaded for the language `#1'. Using the pattern for}%
\typeout{** the default language instead.}%
\else
\language=\csname l@#1\endcsname
\fi
#2}}
\providecommand{\BIBdecl}{\relax}
\BIBdecl

\bibitem{arikan_2009}
E.~{Arikan}, ``{Channel Polarization: A Method for Constructing
  Capacity-Achieving Codes for Symmetric Binary-Input Memoryless Channels},''
  \emph{IEEE Transactions on Information Theory}, vol.~55, no.~7, pp.
  3051--3073, July 2009.

\bibitem{nb_polar_rs}
R.~{Mori} and T.~{Tanaka}, ``{Non-binary Polar Codes Using Reed-Solomon codes
  and Algebraic Geometry Codes},'' in \emph{IEEE Information Theory Workshop,
  2010}, Aug 2010, pp. 1--5.

\bibitem{urbanke_polar_2010}
S.~B. {Korada}, E.~{Sasoglu}, and R.~{Urbanke}, ``{Polar Codes:
  Characterization of Exponent, Bounds, and Constructions},'' \emph{IEEE
  Transactions on Information Theory}, vol.~56, no.~12, pp. 6253--6264, Dec
  2010.

\bibitem{QUP}
K.~{Niu}, K.~{Chen}, and J.~{Lin}, ``{Beyond Turbo Codes: Rate-compatible
  Punctured Polar Codes},'' in \emph{IEEE International Conference on
  Communications (ICC)}, June 2013, pp. 3423--3427.

\bibitem{novel_puncturing}
R.~{Wang} and R.~{Liu}, ``{A Novel Puncturing Scheme for Polar Codes},''
  \emph{IEEE Communications Letters}, vol.~18, pp. 2081--2084, Dec 2014.

\bibitem{shin_length_compatible}
D.~{Shin}, S.~{Lim}, and K.~{Yang}, ``{Design of Length-Compatible Polar Codes
  Based on the Reduction of Polarizing Matrices},'' \emph{IEEE Transactions on
  Communications}, vol.~61, no.~7, pp. 2593--2599, July 2013.

\bibitem{declercq_icc_2017}
L.~{Chandesris}, V.~{Savin}, and D.~{Declercq}, ``{On Puncturing Strategies for
  Polar Codes},'' in \emph{2017 IEEE International Conference on Communications
  Workshops (ICC Workshops)}, May 2017, pp. 766--771.

\bibitem{Storn1997}
\BIBentryALTinterwordspacing
R.~Storn and K.~Price, ``{Differential Evolution -- A Simple and Efficient
  Heuristic for global Optimization over Continuous Spaces},'' \emph{Journal of
  Global Optimization}, vol.~11, no.~4, pp. 341--359, Dec 1997. [Online].
  Available: \url{https://doi.org/10.1023/A:1008202821328}
\BIBentrySTDinterwordspacing

\bibitem{harish_construction}
{Harish Vangala, Emanuele Viterbo, Yi Hong}, ``{A comparative study of polar
  code constructions for the AWGN channel},'' \emph{arxiv.org}, 2015.

\bibitem{trifonov}
P.~{Trifonov}, ``{Efficient Design and Decoding of Polar Codes},'' \emph{IEEE
  Transactions on Communications}, vol.~60, no.~11, November 2012.

\bibitem{5G_NR}
{ Peiying Zhu}, ``{Polar Code for 5G NR},'' ITW 2018 Keynote, available online:
  {http://itw2018.org/static/resource/Keynote-Peiying.pdf}.

\bibitem{list_decoding}
I.~{Tal} and A.~{Vardy}, ``{List Decoding of Polar Codes},'' \emph{IEEE
  Transactions on Information Theory}, vol.~61, no.~5, pp. 2213--2226, May
  2015.

\end{thebibliography}

\end{document}